\documentclass[conference]{IEEEtran}
\usepackage{blindtext, graphicx, todonotes}
\usepackage[utf8]{inputenc}
\ifCLASSINFOpdf
\else
\fi
\usepackage{cclicenses}

\usepackage{xcolor}

\usepackage{hyperref, cite}

\usepackage{flushend}

\IEEEoverridecommandlockouts
\begin{document}
%
\title{Track 1 Paper: Good Usability Practices \\ in Scientific Software Development}

\author{\IEEEauthorblockN{Francisco Queiroz}
\IEEEauthorblockA{Tecgraf Institute, PUC-Rio\\
  Department of Arts \& Design, PUC-Rio\\
  Rio de Janeiro, Brazil}
\\
\IEEEauthorblockN{Sandor Brockhauser}
\IEEEauthorblockA{
  European XFEL GmbH\\
  Schenefeld, Germany}
\and
\IEEEauthorblockN{Raniere Silva}
\IEEEauthorblockA{Software Sustainability Institute UK\\
  School of Computer Science\\
  University of Manchester\\
  Manchester, United Kingdom}
\and
\IEEEauthorblockN{Jonah Miller}
\IEEEauthorblockA{Perimeter Institute of Theoretical Physics\\
  Waterloo, Canada}
\IEEEauthorblockA{University of Guelph\\
  Guelph, Canada}
\\
\IEEEauthorblockN{Hans Fangohr}
\IEEEauthorblockA{University of Southampton\\
  United Kingdom\\
  European XFEL GmbH\\
  Schenefeld, Germany}

\thanks{Licensed under a \href{https://creativecommons.org/licenses/by/4.0/}{CC-BY-4.0 license}. DOI: 10.6084/m9.figshare.5331814.}

}


%


\maketitle

\begin{abstract}
Scientific software often presents very particular requirements regarding 
usability, which is often completely overlooked in this setting.
As computational science has emerged as its own discipline, distinct from theoretical and experimental science, it has put new requirements on future scientific software developments. 
In this paper, we discuss the background of these problems and introduce nine 
aspects of good usability. We also highlight best practices for each aspect 
with an emphasis on applications in computational science. 
\end{abstract}


%
\IEEEpeerreviewmaketitle

\section{Introduction}
Scientific software development is a field of growing importance
but lacks a widespread methodology.
Scientists generally have little or no training
in software engineering but tend to be main developers of
computational science codes. They face a number of challenges including:
quickly changing requirements due to the research nature of the work,
competition between maintainable and performance code, and lack of metrics
that would reward investment into sustainable software~\cite{Segal:2007, Kelly:2007}. 
Of particular detriment is the pressure to rapidly produce scientific
publications~\cite{Wilson:2006, Killcoyne:2009}. 
It may be possible to overcome this publication pressure 
when funding agencies are convinced that it is 
worth investing directly in software software for computationally intensive fields. 
The Science and Technology Facilities Council (STFC) in the UK Collaborative 
Computational Projects (\url{http://www.ccp.ac.uk/about.html}) 
sets a good example.

In this work we focus on \emph{usability}, a particular aspect of
software development and design. Usability  is one of the attributes of
sustainable software and can be defined as
``the extent to which a product can be used by specified users to achieve
specified goals with effectiveness, efficiency, and satisfaction in a specified
context of us''~\cite[p.3]{Venters_WSSSPE}. Without proper usability, a software 
cannot be distributed and applied even within its targeted domain.
More importantly, its unusable software can easily result in non-reproducible 
science and the violation of the FAIR principles~\cite{Wilkinson:2016}. 
Unfortunately, usability is often neglected in scientific software development~\cite{Ahmed:2014},
and is of mixed perceived importance to users and developers~\cite{Nguyen-Hoan:2010, Hucka:2016}.
Scientific software usage and development present many challenges for usability 
design that can be related to development models, user-base needs and 
specialization, professional practices, technical constraints, and scientific 
demands~\cite{Queiroz:2016}. Computational science is therefore 
an idiosyncratic field with unique and, occasionally, counterintuitive usability 
requirements. There are, nevertheless, a significant number of informative 
case studies and guidelines on the subject~\cite{MacLeod:1992, Springmeyer:1993, 
Pancake:1996, Javahery:2004, Schraefel:2004,Letondal:2004,Talbott:2005, 
Macaulay:2009, DeRoure:2009, Keefe:2010, DeMatos:2013,Ahmed:2014, Fangohr:2016, 
Beg:2016,List:2017}. Supported by those references and informed by first-hand
experience, we discuss usability challenges and how to address them.

\section{Good Practices}

\subsection{Think Beyond Graphical User Interfaces}
\label{sec:beyond:GUIs}
Graphical user interfaces (GUIs) have made software user-friendly and 
arguably fosteed the popularization of software in general. However, scientific 
software might require alternatives that, if not more intuitive, are more 
appropriate and efficient depending on the user's needs --- especially if they 
involve entering a large amounts of data, and reading the data from many files, 
or running on a shared or distributed architectures.
Command-Line Interfaces (CLIs) are popular in computational science because they 
often allow for quick repetition of tasks~\cite{bestprSC} and scriptability.
The analysis of large datasets can be significantly easier and more productive 
when done through command-line input than through visual-based interfaces~\cite{Springmeyer:1993}.\footnote{Of course using a CLI does not guarantee ease of use. One must still follow usability best practices when designing the CLI.}
Moreover, GUIs can be extremely cumbersome on distributed infrastructures 
such as supercomputers.
To implement a GUI for distributed code, a graphical frontend must connect via 
network to a distributed backend. Although many scientific visualization tools 
such as VisIt~\cite{HPV:VisIt} and Paraview~\cite{ahrens2005paraview, ayachit2015paraview} 
have implemented this scheme, full rendering via GUI can be impractical and computationally intensive renders are often performed ``headless'' without the GUI~\cite{HPV:VisIt}.
Because of these difficulties, most distributed scientific codes completely 
lack a graphical frontend or separate computation and visualization into separate and 
subsequent stages in the workflow. Similarly, complex experimental protocols 
combining data analysis and scientific instrumentation control can be designed 
with separate User Interaction points in the complete workflow~\cite{Brockhauser:gm5021}.

Even for software where daily use relies on a GUI, such as text
processors and web browsers, there are times when having a CLI for
some tasks is a time saver. For example, users of LaTeX, or the
LibreOffice or Chrome CLIs can convert a text document into PDF format
from the command line.

\subsection{Keep UI Code Separate From Scientific Calculation}

Simulation (or any scientific calculation) should not be embedded in User Interface (UI) code~\cite{Kelly:2009}.
This rule is particularly true for scientific software primarily because, as previously stated, scientific software should be usable via a number of alternative interfaces, such as GUI and CLI. Moreover, it should be possible to access these interfaces both locally and over a network (e.g., via ssh).
Keeping the scientific calculation code wrapped into functions that are called by the UI should
make reconfiguration and customization more convenient~\cite{Bastos:2013},
make porting the functionalities to another UI easier
and
make integration with other software simpler.


\subsection{Keep the Configuration in a File} \label{sec:beyond:GUIs}

Some tasks requires researchers to provide a long list of parameters to define their computational problem, 
and the software they are using may not provide default values for the parameters, or the default parameters need to be overriden.
In these situations, it
is very handy to have the ability to store some or all of the parameters
in a configuration file that the software can read
at the begining of every execution. 
Alternatively, if a command line tool asks for input parameters from 
the standard input, which requires continuous user interaction, 
it can be modified to be scriptable.
In a script, the configuration parameters are stored next to the 
execution command itself~\cite{Potterton:hv0002}.

Configuration files have the advantage of being \textit{declarative} and
\textit{automatically verifiable}. 
The file defines a \textit{state} which the program will start from or try to achieve, rather than a procedure which leads to that state. Moreover, a parser can automatically check to see if state is valid.
The former is good for reproducibility because it is (ideally) unambiguous even years later~\cite{DBLP:conf/agp/Lloyd94}. 
The latter is good for accuracy because the code can check if the parameters in the file are sensible~\cite{Beg:2016}. 
This state-based approach can also make parallelism easier to automatically reason about~\cite{chakravarty1997massively} and some parallel runtime environments have made use of this property~\cite{Buss:2010:SST:1815695.1815713,Bauer:2012:LEL:2388996.2389086}.

Domain Specific Languages (DSLs), on the other hand,
provide additional flexibility not present in a plain configuration file.
They allow the user to \textit{programmatically} define new behaviour for the code.
This can be a major advantage since it often enables the code to be extended to
unforeseen use-cases without major rewrites.
DSLs can also allow the user to interact with the code at runtime, which can be
helpful for debugging, prototyping, and visualization~\cite{Beg:2016}.
The syntax and rules of the DSL can also provide the same error checking as a
parameter file.

While defining a new domain specific language (DSL) requires
the development of a parser for the language, 
this extra effort can be avoided by
\emph{embedding} the domain specific language in an existing general
purposes language. This has been demonstrated by a number of projects
recently, and Python is a common choice as the general purpose
language. In this context, the domain specific language is given
through a Python module that the user imports into their generic
Python program, and which provides commands, objects and operations
that are specific to the domain in question. The Python program 
then becomes the (very flexible) configuration file for the computational problem.

However, some care is required when designing such DSLs: the elements
of the DSL must be constructed so that users cannot combine them in
ways that would take the tool outside its range of applicability. This
could be achieved through explicit assert statements in the DSL's
implementation or appropriate (often Object Oriented) design. If the
code author lacks the experience or time available to achieve this, it
is important to document the assumptions made for use of the DSL so
that it is not used incorrectly inadvertently by others in the future. For 
example, the yt project~\cite{2011ApJS..192....9T} is a DSL for scientific 
visualization and data analysis built in Python. If the data fed into yt 
doesn't satisfy the correct assumptions, yt could produce spurious 
visualization artifacts or incorrectly integrate a quantity over the domain. 
Therefore the authors of yt take extreme care to document their API, 
sanitize their inputs and throw informative error messages when incorrect 
data is fed into the tool. A major part of this process is unit testing the 
DSL's functionality

Some codes combine both plain configuration files \textit{and} DSLs.
For example, the Einstein Toolkit~\cite{Loffler:2011ay},\footnote{For which one 
of us is a developer.} a code for relativistic astrophysics, uses configuration 
files for day-to-day simulations. However, it also provides a low-level DSL, 
Kranc~\cite{Husa:2004ip}, for defining systems of equations to solve.

\subsection{Design for Small, Incremental Changes}

Making incremental changes is considered a best practice for scientific software 
development~\cite{bestprSC}, and the same principle applies to user interfaces. 
Ideally, \emph{UIs should be planned for extensibility and frequent changes} as 
new requisites emerge. Through incremental changes, software is more likely to 
\emph{stay attuned to users’ needs}, not forcing them to radically change the 
way they work~\cite{DeRoure:2009}. 

Regarding constant updates and addition of 
new functionalities, UI components that can be easily extended might offer 
interesting solutions. This is the case for map3D, a scientific 
visualization software for displaying and editing three-dimensional models and
associated data~\cite{SCI:Map3d}. During development, pop-up menus were
implemented for providing the necessary flexibility, allowing 
developers to add new commands and submenus as the software development 
and requisites evolved~\cite{MacLeod:1992}. It is worth mentioning that
web-based applications might take advantage of the modularity allowed by 
frontend design methodologies such as Atomic Design~\cite{Atomic}, making
it easier to configure user interfaces as the project advances.

The parameter files and DSLs described in section~\ref{sec:beyond:GUIs}
are particularly good for satisfying this design
constraint. For example, the Einstein Toolkit~\cite{Loffler:2011ay}
packages low-level code in modules. Each module must declare which
functionality it adds, which relevant parameters can be set in the
parameter file, and how these parameters depend on other modules. The parameter
file parser then automatically adds these options to the parameter
file at compile time. The yt project~\cite{2011ApJS..192....9T}
provides a DSL for scientific visualization. This DSL interacts with
the low-level code only through function calls and so functionality
can easily be incrementally added by the introduction of new DSL
language features or functions.

\subsection{Facilitate and Register User Activity and Environment}
There are a number of ways through which usability can be enhanced based on \emph{past} user activity.
First, providing access to a list of recent commands and allowing
users to re-execute them can help users save time.
This is a major reason for the popularity of command-line interfaces~\cite{bestprSC}.
A very popular implementation of this concept is the ability to access previously typed
commands by pressing the up arrow key or do a reverse search on the history of executed commands.
Users can also press the right and left arrow keys to navigate through a previous command
and edit it to suit their needs.
Second, it might be a good idea to give users quick access
to frequently used commands~\cite{Julvez:2014}.
In some environments, the tab key is used to roll among frequent used commands
or to auto complete a command.
Third, logging user activity
might help users identify and support research reproducibility~\cite{List:2017}
by exporting the history to a file.

After registering user activity,
developers can go further and log the user environment,
i.e. compiled binary, configuration files, input files and output files,
used when running the program.
This is useful in scientific software since the output of any experiment can be different because of different implementations (or compiler optimizations) of
the Basic Linear Algebra Subprograms (BLAS), LAPACK (Linear Algebra Package) or
any other library used when performing the experiment.
This automatic logging not only provides users
quick access to their exact configuration for debugging purposes
but also allows the computation to be reproduced years after it was run for the first time.
One example framework is Formaline~\cite{formaline}.
For software developers working with Python, we
mention the related packages ReciPy~\cite{ReciPy} and Sumatra~\cite{sumatra2014}.

\subsection{Learn About How Users Work}

Guidelines and case studies often recommend the adoption of a user-centered 
design process that seeks to develop a firm understanding of how scientists do 
their work before developing a piece of software.
This understanding can be acquired by learning the meanderings of scientific work~\cite{Springmeyer:1993}, 
or through a participatory design approach in which users are actively involved in 
the design process~\cite{Letondal:2004,Aragon:2008, Thomer:2016,  Luna2017204}.
It is also important to \emph{analyze the scientific work within the 
environment where it actually takes place}~\cite{Pancake:1996} and evaluate 
existing tools which are already in use~\cite{Javahery:2004}. In this last case
it might be advantageous to adopt preexisting industry standards (e.g.: keyboard
shortcuts for common functionalities, iconography, etc.).  

When designing user interfaces for scientific software, it is a good idea to 
\emph{address specific users or user-bases} rather than aim for a general 
solution~\cite {Javahery:2004, DeRoure:2009}. Ideally, GUIs should be open to 
user customization and adjustable to personal preferences and professional 
specialization~\cite{Gertz:1994, Javahery:2004}. However, users should not be
overwhelmed by an excessive number of customizable parameters --- some 
of which can be unimportant or meaningless to their specific case.
Instead, there should be an additional section for setting advanced 
parameters~\cite{List:2017}. 

As a user base grows, users may have suggestions for improving the UI
or the underlying scientific code. If the code is open source, it can
be extremely advantageous to transform these \textit{users} into
\textit{developers} so that they can bring their user experience and
domain expertise to bear~\cite{Turk:2013:SCH:2484762.2484782}.
Additionally, it is advisable that scientific domain experts are brought into 
the design process for informing \emph{domain best 
practices}~\cite{Schraefel:2004,  DeMatos:2013} and evaluating the 
tool~\cite{Keefe:2010}.

\subsection{Be Minimalistic, but Look Out for Exceptional Needs}

Designers should be attentive to information that is particularly relevant in
scientific software, but that could be eluded otherwise. \emph{Metadata},
for instance, is often required to be readable and easy to
access~\cite{Talbott:2005, Baxter:2006, Macaulay:2009, Keefe:2010, bestprSC, 
Thomer:2016}. 

Also, despite recent trends
favoring flat design over skeuomorphism (i.e.: visual design that imitates the 
appearance of real-world objects), software versions of physical
instruments might benefit from adopting the looks of their real-world
counterparts~\cite{Foster:1998}, making it easier for users to recognize and 
learn about their functioning. An example for that approach is LabViEW's 
set of GUI components mimicking dials, knobs and meters~\cite{LabVIEW}.

Finally, minimalism should \emph{emphasize, rather than
conceal, critical information} such as system malfunctioning~\cite{Morais:2014},
emergency information~\cite{Ferguson:2016}, and situations where awareness and
response under time pressure are essential. For instance, the
Sky software for astronomical visualization reduces users' cognitive load by 
simplifying three-dimensional visualization data as a two-dimensional 
projection~\cite{Aragon:2008}.

\subsection{Design for Precision}

In order to achieve satisfying results, scientific work often demands precision
regarding user's input. A possible means for that would be continuously 
constraining users' input and providing feedback on it. The Dynamic Dragging
Interface, for instance, makes user of force-feedback input devices to 
help users selecting sections of 3D brain visualization ~\cite{Keefe:2010}.
Another solution would be not accepting the input when 
the input device, such as a stylus or mouse, moves too fast. 

It can be advantageous to have two input modes for the same action ---
one designed for accuracy,
and another for speed. In map3D, for instance, geometric models could be moved, 
rotated and scaled through dial boxes (accurate, but slower) or, alternatively, 
via mouse (less accurate, but faster)~\cite{MacLeod:1992}. Another approach 
would be to give users a way to switch between fast and precise working modes.
For instance, by activating a 'snap' mode where the mouse cursor snaps to a gridline,
objects or other elements on screen. 


\subsection{Contextualize User Actions}
Work in scientific software can involve a number of different and/or sequential
tasks to be performed by users. In those cases, it is desirable to contextualize
users' actions, facilitating their access to functions that are relevant to their 
current tasks and preventing their access to functions that are not.

The Petri Net Toolbox for MATLAB, for instance, features a button that toggles 
between \emph{Draw Mode}, for creating and editing models, and \emph{Explore Mode},
for simulation and analysis~\cite{Julvez:2014}. When switching between modes, 
GUI elements are displayed or hidden depending on their relevance to the 
selected mode. This approach is known as a design pattern 
named  \emph{Disabled Irrelevant Things}~\cite{Zeeshan:2011}. 

Another possible approach is the \emph{Window Per Task}~\cite{Zeeshan:2011} design pattern, in which tasks are distributed
across individual screens containing the appropriate commands for that task only. Pharmaceutical
biology software Lipid-Pro makes use of this design pattern by organizing its tasks into separate
panels~\cite{Ahmed:2014}.

\section{Summary}
Throughout the previous section, we have presented a nonexhaustive set of good 
practices in usability for scientific software, taking in consideration 
challenging aspects of scientific software development and use such as the lack 
of attention to software engineering; the need for reproducibility; the handling 
of large amounts of data; the complexity of actions and parameters involved in 
scientific work; frequent changes in requirements; particularities of scientific 
work and its environment; the need for accessing and responding to critical 
information; and the importance of precision.

By adopting the presented practices, developers should be
able to deliver applications that are more usable, robust and more appropriate 
for scientific work.


%

\section*{Acknowledgment}

This project began during the Fourth Annual Workshop on Sustainable
Software for Science: Practices and Experiences (WSSSPE4), where four of the 
authors established the Working Group on Software best practices for 
undergraduates~\cite{WSSSPE4Report}. The authors would therefore like to thank the
organizers of WSSSPE4 for facilitating this conversation. J. Miller
and F. Queiroz would like to thank travel grants from the 
National Science Foundation of the USA and The  Gordon  and  Betty  Moore  
Foundation, which made attendance possible. J. Miller also acknowledges 
support from the Natural Sciences and Engineering Research Council of 
Canada and from the National Science Foundation of the USA (OCI 0905046, 
PHY 1212401). Research at Perimeter Institute is supported by the Government of
Canada through the Department of Innovation, Science and Economic
Development and by the Province of Ontario through the Ministry of
Research and Innovation. F. Queiroz acknowledges support from PUC-Rio and 
Tecgraf Institute.
H. Fangohr and R. Silva acknowledge support from the Software Sustainability Institute
and Engineering and Physical Sciences Research Council (EPSRC) in the UK.


\ifCLASSOPTIONcaptionsoff
  \newpage
\fi



%
\bibliographystyle{IEEEtran}

\bibliography{main}
%





\end{document}